\documentclass[11pt,titlepage,a4paper]{article}
\usepackage{bozhomac}	% automatically AMS LaTeX package
\usepackage{bozhlogo}
\usepackage{xr1}
\title{\bfseries	\vspace*{-2.345in}
\vspace*{-2.6ex}
{
\begin{flushright}
	  \textbf{\large LANL xxx E-print archive No. quant-ph/9804062}\\[2ex]
\end{flushright}
}
{\huge Fibre bundle formulation of	\\[6pt]
nonrelativistic quantum mechanics	\\[2ex]
\Large II. Equations of motion and observables
}
}

\vspace{3ex}

\author{
Bozhidar Z. Iliev
\thanks{Department Mathematical Modeling,
Institute for Nuclear Research and \mbox{Nuclear} Energy,
Bulgarian Academy of Sciences,
Boul. Tzarigradsko chauss\'ee~72, 1784 Sofia, Bulgaria}
\thanks{E-mail address: bozho@inrne.bas.bg}
\thanks{URL: http://www.inrne.bas.bg/mathmod/bozhome/}
}

\date{
\vspace{2.27ex}\ShortTitle{Bundle quantum mechanics: II}	\\[0.27ex]
\vspace{3.27ex}
	\begin{tabular}{r@{$\colon\to~$}l}
\vspace{0.09ex} Basic ideas 	& March 1996		\\[0.09ex]
\vspace{0.09ex} Began		& May 19, 1996		\\[0.09ex]
\vspace{0.09ex} Ended		& July 12, 1996		\\[0.09ex]
\vspace{0.09ex} Revised		& December 1996 -- January 1997,\\[0.09ex]
\vspace{0.09ex} Revised		& April 1997, September 1998	\\[0.09ex]
\vspace{0.09ex} Last update	& October 21, 1998  	\\[0.09ex]
\vspace{0.09ex} Composing/Extracting part II
			& September 27/October 4, 1997	\\[0.09ex]
\vspace{0.09ex} Updating part II   & October 21, 1998 	\\[0.09ex]
\vspace{0.27ex} Produced	   & \fbox{\today}	\\[0.27ex]
	\end{tabular} \\[1.27ex]
	\begin{tabular}{r@{$\colon~$}l}
\vspace{0.27ex} LANL xxx archive server E-print No. &	quant-ph/9804062
							 \\[0.27ex]
	\end{tabular} \\[-0.27ex]
\vspace{3.27ex}{\Huge\BOZHO}	\\[3.27ex]
\vspace{0.27ex}\Subject{Quantum mechanics; Differential geometry} \\[2.27ex]
	\begin{tabular}{r@{\hspace{0.512em}}|@{\hspace{0.512em}}l}
\vspace{0.27ex}		  % \\[0.27ex]
&
\vspace{0.27ex}% \\[0.27ex]
	\end{tabular} \\[1.27ex]
\vspace{0.27ex}	\\[0.27ex]
}
\newcommand{\Hil}{\mathcal{F}}	% usual Hilbert space ----->>
\newcommand{\HilB}{(\bHil,\proj,\base)}	% Hilbert fibre bundle
	\newcommand{\bHil}{\mathit{F}}	% (total) bundle Hilbert space
	\newcommand{\proj}{\pi}		% Hilbert bundle projection
	\newcommand{\base}{\mathit{M}}	% Hilbert bundle base space

\newcommand{\Ham}{\mathcal{H}}	% usual Hamiltonian
\newcommand{\bHam}{\mathit{H}}	% bundle Hamiltonian morphism

\newcommand{\HamM}{\boldsymbol{\Ham}} % the matrix of the usual Hamiltonian
\newcommand{\bHamM}{\boldsymbol{\bHam}} % the matrix of the bundle Hamiltonian

\newcommand{\mHam}{\boldsymbol{\Ham}^\mathbf{m}}   % usual matrix Hamiltonian
\newcommand{\mbHam}{\boldsymbol{\bHam}\!^\mathbf{m}}%bundle matrix Hamiltonian

\newcommand{\dyn}[1]{\pmb{\mathbb{#1}}}	% dynamical variable ----->>
	\newcommand{\ope}[1]{\mathcal{#1}}		 % operator
	\newcommand{\mope}[1]{\boldsymbol{\mathcal{#1}}} % matrix of operator
	\newcommand{\mor}[1]{\mathit{#1}}		 % morphism
	\newcommand{\mmor}[1]{\boldsymbol{\mathit{#1}}}	 % matrix of morphism

\newcommand{\ih}{\mathrm{i}\hbar}% imaginary unit times Plank constant/{2\pi}
\newcommand{\iih}{\frac{1}{\ih}} % the inverse to \ih (see previous command)

\begin{document}		% BEGINNING OF THE DOCUMENT

% 		FIXING REFERENCES TO PREVIOUS PART(S)

% 		references to part I
\renewcommand*{\thesection}{I.\arabic{section}}
\renewcommand*{\thefootnote}{\protect{I.\arabic{footnote}}}
\setcounter{page}{0}
\setcounter{footnote}{0}
\setcounter{section}{0}
\setcounter{equation}{0}
 	\include{bqm-1txt}
\renewcommand*{\thesection}{\arabic{section}}
\renewcommand*{\thefootnote}{\arabic{footnote}}
\setcounter{page}{0}
\setcounter{footnote}{0}
\setcounter{section}{0}
\setcounter{equation}{0}
% 		end of references to part I

% 		END OF FIXING REFERENCES TO PREVIOUS PART(S)

\renewcommand{\thefootnote}{\fnsymbol{footnote}}
\maketitle			% the title (page) is put here
\renewcommand{\thefootnote}{\arabic{footnote}}

\tableofcontents		% the table of contents is being put here

%%%%%%%%%%%%%%%%%%%%%%%%%%%%%%%%%%%%%%%%%%%%%%%%%%%%%%%%%%%%%%%%%%%%%%%%%%%%%
%%%%%									%%%%%
%%%%%		actual beginning of the document			%%%%%
%%%%%									%%%%%
%%%%%%%%%%%%%%%%%%%%%%%%%%%%%%%%%%%%%%%%%%%%%%%%%%%%%%%%%%%%%%%%%%%%%%%%%%%%%

\pagestyle{myheadings}
\markright{\itshape\bfseries Bozhidar Z. Iliev:
	\upshape\sffamily\bfseries Bundle quantum mechanics.~II}

\begin{abstract}

We propose a new systematic fibre bundle formulation of nonrelativistic
quantum mechanics. The new form of the theory is equivalent to the usual one
but it is in harmony with the modern trends in theoretical physics and
potentially admits new generalizations in different directions. In it
a pure state of some quantum system is described by a state section (along
paths) of a (Hilbert) fibre bundle. It's evolution is determined through the
bundle (analogue of the) Schr\"odinger equation. Now the dynamical variables
and the density operator are described via bundle morphisms (along paths).
The mentioned quantities are connected by a number of relations derived in
this work.
%
% The bundle analogues of the different pictures of motion are
% investigated too.
%

	In the second part of this investigation we derive several forms of
the bundle (analogue of the) Schr\"odinger equation governing the time
evolution of state sections. We prove that up to a constant the
matrix-bundle Hamiltonian, entering in the bundle analogue of the matrix
form of the conventional Schr\"odinger equation, coincides with the matrix
of coefficients of the evolution transport. This allows to interpret the
Hamiltonian as a gauge field. Here we also apply the bundle approach to the
description of observables. It is shown that to any observable there
corresponds a unique Hermitian bundle morphism (along paths) and vice versa.

\end{abstract}

\vspace*{-8.5ex}
\section {Introduction}
\label{introduction-II}
\setcounter{equation} {0}

	This paper is a second part of our investigation devoted to the fibre
bundle approach to nonrelativistic quantum mechanics. It is a straightforward
continuation of~\cite{bp-BQM-introduction+transport}.

	The organization of the material is the following.

	Sect.~\ref{V} is devoted to the bundle analogues of the Schr\"odinger
equation which are fully equivalent to it. In particular, in it is introduced
the
\emph{matrix-bundle Hamiltonian}
which governs the quantum evolution through the
\emph{matrix-bundle Schr\"odinger equation}.
The corresponding matrix of the bundle-evolution transport (operator) is
found. It is proved that
\emph{up to a constant the matrix of the coefficients of the bundle
evolution transport coincides with the matrix-bundle Hamiltonian}.
On this basis is derived the
\emph{(invariant) bundle-Schr\"odinger equation}.
Geometrically it simply means that the corresponding state sections are
(parallelly, or, more precisely, linearly) transported by means of the
bundle evolution transport (along paths).

	In Sect.~\ref{VI} is considered the question for the bundle
description of observables. It turns out that
\emph{to any observable there corresponds a unique Hermitian bundle
morphism (along paths) and vice versa}.

	Sect.~\ref{conclusion-II} closes the work.
\vspace{1.2ex}

	The notation of the present work is the the same as the one
in~\cite{bp-BQM-introduction+transport}
and we are not going to recall it here.

	The references to sections, equations, footnotes etc.
from~\cite{bp-BQM-introduction+transport} are obtained from their sequential
numbers in~\cite{bp-BQM-introduction+transport} by adding in front of them the
Roman one (I) and a dot as a separator. For instance, Sect.~I.4 and (I.5.13)
mean respectively section 4 and equation~(5.13) (equation 13 in Sect.~5)
of~\cite{bp-BQM-introduction+transport}.

	Below, for reference purposes, we present a list of some essential
equations of~\cite{bp-BQM-introduction+transport} which are used in this paper.
Following the just given convention, we retain their original reference
numbers.
\enlargethispage*{5.5ex}
	\begin{gather}%++++++++++++++++++++++++++++++++++++++++++++++++++++
\tag{\ref{2.5}}% \label{2.5}
\ih\frac{d \psi(t)}{d t} = \Ham(t) \psi(t) ,
\\
\tag{\ref{2.6}}% \label{2.6}
\ih\frac{\partial \ope{U}(t,t_0)}{\partial t} = \Ham(t)\circ\ope{U}(t,t_0),
\qquad
\ope{U}(t_0,t_0) = \id_\Hil ,
\\
\tag{\ref{2.7}}% \label{2.7}
\Ham(t) =
\ih\frac{\partial \ope{U}(t,t_0)}{\partial t} \circ \ope{U}^{-1}(t,t_0) =
\ih\frac{\partial \ope{U}(t,t_0)}{\partial t} \circ \ope{U}(t_0,t) ,
\\
\tag{\ref{2.9}}% \label{2.9}
\langle\ope{A}(t)\rangle_\psi^t :=
\frac{\langle\psi(t) | \ope{A}(t)\psi(t)\rangle}
{\langle\psi(t) | \psi(t)\rangle} ,
\\
\tag{\ref{4.3}}% \label{4.3}
\psi(t) = l_{\gamma(t)} ( {\Psi}_\gamma(t) ) \in \Hil ,
\\
\tag{\ref{4.8}}% \label{4.8}
\langle \cdot|\cdot \rangle _x=
\langle l_{x}\cdot|l_{x}\cdot \rangle, \qquad x\in \base ,
\\
\tag{\ref{4.morphism}}% \label{4.morphism}
\langle \mor{A}^\ddag\Phi_x | \Psi_x \rangle_x :=
\langle \Phi_x | \mor{A}\Psi_x \rangle_x,
\qquad \Phi_x,\Psi_x\in \bHil_x ,
\\
\tag{\ref{4.4}}% \label{4.4}
\Psi_\gamma(t) = \mor{U}_\gamma(t,s) \Psi_\gamma(s) ,
\\
%	\end{gather}%++++++++++++++++++++++++++++++++++++++++++++++++++++++
%	\begin{gather}%++++++++++++++++++++++++++++++++++++++++++++++++++++
%
\tag{\ref{4.7}}% \label{4.7}
\mor{U}_\gamma(t,s) =
		l_{\gamma(t)}^{-1}\circ \ope{U}(t,s) \circ l_{\gamma(s)},
\qquad s,t\in J .
	\end{gather}%++++++++++++++++++++++++++++++++++++++++++++++++++++++
\newpage

\section {The bundle equations of motion}
\label{V}
\setcounter{equation} {0}

	If we substitute~(\ref{4.7'}) into~(\ref{2.5})--(\ref{2.8}),
we `get' the `bundle' analogues of~(\ref{2.5})--(\ref{2.8}).
But they will be wrong! The reason for this being that they will contain
partial derivatives like
\(
\partial l_{\gamma(t)}/\partial t,\
\partial \Psi_\gamma(t)/\partial t,\) and
\(
\partial \mor{U}_\gamma(t,t_0)/\partial t,
\)
which are not defined at all. For instance, for the first one we must have
\(
\partial l_{\gamma(t)}/\partial t =
\lim_{\varepsilon\to0}\left(	\frac{1}{\varepsilon}
( l_{\gamma(t+\varepsilon)} - l_{\gamma(t)} )	\right),
\)
but the `difference' in this limit is not defined (for $\varepsilon\not=0$)
because $l_{\gamma(t+\varepsilon)}$ and $l_{\gamma(t)}$ act on different
spaces, viz.\ resp.\  on
$\bHil_{\gamma(t+\varepsilon)}$ and $\bHil_{\gamma(t)}$.
The same is the situation with $\partial \mor{U}_\gamma(t,t_0)/\partial t$.
The most obvious is the contradiction in
\(
\partial \Psi_\gamma(t)/\partial t =
\lim_{\varepsilon\to0}\left(	\frac{1}{\varepsilon}
( \Psi_\gamma(t+\varepsilon) - \Psi_\gamma(t) )\right),
\)
because $\Psi_\gamma(t+\varepsilon)$ and $\Psi_\gamma(t)$ belong to different
(for $\varepsilon\neq0$) vector spaces.

	One way to go through this difficulty is to define, e.g.
$\partial \Psi_\gamma(t)/\partial t$ like
$l_{\gamma(t)}^{-1}\partial \psi_\gamma(t)/\partial t$
(cf.~(\ref{4.3})) but this does not lead to something important and new.

	To overcome this problem, we are going to introduce local bases (or
coordinates) and to work with the matrices of the corresponding
operators and vectors in them.

	Let $\{e_a(x),\quad a\in\Lambda\}$ be a basis in
$\bHil_x=\proj^{-1}(x),\ x\in \base$.
The indices $a,b,c,\ldots\in\Lambda$ may
take discrete, or continuous, or both values. More precisely, the set
$\Lambda$ has a decomposition $\Lambda=\Lambda_d\bigcup\Lambda_c$ where
$\Lambda_d$ is a union of (a finite or countable) subsets of
$\mathbb{N}$ (or, equivalently, of
$\mathbb{Z}$) and $\Lambda_c$ is union of subsets of $\mathbb{R}$ (or,
equivalently, of $\mathbb{C}$). Note that  $\Lambda_d$ or  $\Lambda_c$,
but not both, can be empty.
	This is why sums like $\lambda^a e_a(x)$ or $\lambda_a\mu^a$ for
$a\in\Lambda,\ \lambda^a,\mu_a\in\mathbb{C}$ must be understood as a sum over
the discrete (enumerable) part(s) of $\Lambda$, if any, plus the
(Stieltjes or Lebesgue) integrals over the continuous part(s) of $\Lambda$,
if any. For instance:
\(
\lambda^ae_a(x):=\sum_{a\in\Lambda} \lambda^ae_a(x) :=
\sum_{a\in\Lambda_d} \lambda^ae_a(x) +
\int_{a\in\Lambda_c} \lambda^ae_a(x) da.
\)
	By this reason it is better to write
\(
\int \!\!\!  \!\!\!  \sum_{a\in\Lambda}:=
\sum_{a\in\Lambda_d} + \int_{a\in\Lambda_c} da
\)
	instead of \( \sum_{a\in\Lambda}  \),
but we shall avoid this complicated notation by using the assumed
summation convention on indices repeated on different levels.%
\footnote{%
For details about infinite dimensional matrices see, for
instance,~\cite{Neumann-MFQM} and~\cite[chapter~VII, \S~18]{Messiah-QM}.
A comprehensive presentation of the theory of infinite matrices is given
in~\cite{Cooke-infty-mat}; this book is mainly devoted to infinite discrete
matrices but it contains also some results on continuous infinite matrices
related to Hilbert spaces.%
}

	The matrices corresponding to vectors or operators in a given
field of bases will be denoted with the same symbol but in
\textbf{boldface}, for example:
$\mmor{U}_\gamma(t,s) :=
\left[ \left( \mor{U}_\gamma(t,s) \right)_{{\ }b}^{a}  \right]$
and
$\boldsymbol{\Psi}_\gamma(s) :=
\left[ \Psi_\gamma^{a}(s)  \right]$, where
\(
\mor{U}_\gamma(t,s)\left(e_b(\gamma(s))\right) =:
\left( \mor{U}_\gamma(t,s) \right)_{{\ }b}^{a} e_a(\gamma(t))
\)
and
\(
\Psi_\gamma(s) =: \Psi_\gamma^{a}(s) e_a(\gamma(s)).
\)%
\footnote{%
The matrices $\mope{U}(t,s)$ and $\mmor{U}(t,s)$ are closely related to
propagator functions~\cite{Bjorken&Drell-1}, but we will not need these
explicit connections.  For explicit calculations and construction of
$mope(t,s)$ see~\cite[\S~21, \S22]{Bjorken&Drell-1}%
}

	Analogously, we suppose in $\Hil$ to be fixed a basis
$\{f_a(t),\quad a\in\Lambda\}$ with respect to which we shall use the same
bold-faced matrix notation, for instance:
$\mope{U}(t,s)=\left[ \ope{U}_{{\ }a}^{b}(t,s) \right]$,
\(
\ope{U}(t,s)\left(f_a(s)\right) =:
\left( \ope{U}(t,s) \right)_{{\ }a}^{b} f_b(t)
\),
\(
\boldsymbol{\psi}(t) = \left[ \psi^a(t) \right],\
\psi(t) =: \psi^a(t) f_a(t),
\)
and, at last,
\(
\boldsymbol{l}_x(t) = \bigl[ \left(l_x\right)_{{\ }a}^{b}\!(t)  \bigr],\
l_x\left(e_a(x)\right) =: \left(l_x\right)_{{\ }a}^{b}\!(t) f_b(t).
\)
	Generally $\boldsymbol{l}_x(t)$ depends on $x$ and $t$, but if
$x=\gamma(s)$ for some $s\in J$, we put $t=s$ as from physical reasons
is clear that $\bHil_{\gamma(t)}$ corresponds to $\Hil$ at the `moment'
$t$, i.e.\  the components of $l_{\gamma(s)}$ are with respect to
$\{e_a(\gamma(s))\}$ and
$\{f_a(s)\}$. The same remark concerns `two-point' objects like
$\mor{U}_\gamma(t,s)$
and $\ope{U}(t,s)$
whose components will be taken with respect to pairs of bases like
$( \{e_a(\gamma(t))\} , \{e_a(\gamma(s))\} )$
and
$( \{f_a(t)\} , \{f_a(s)\} )$ respectively.

	Evidently, the equations~(\ref{4.3}), (\ref{4.4})--(\ref{4.7})
remain valid \emph{mutatis mutandis} in the introduced matrix notation:
the kernel letters have to be made bold-faced, the operator composition
(product) must be replaced by matrix multiplication, and the identity
map $\id_{\bHil_x}$ has
to be replaced by the unit matrix
\(
\openone_{\bHil_x} := \left[ \delta_{a}^{b} \right] :=
\left[ \left(\id_{\bHil_x}\right)_{{\ }a}^{b} \right]
\)
of $\bHil_x$ in $\{e_a(x)\}$. Here
$\delta_{a}^{b}=1$ for $a=b$ and
$\delta_{a}^{b}=0$ for $a\neq b$, which means that
\(
e_a(x) = \delta_{a}^{b} e_b(x).
\)
	For instance, using the above definitions, one verifies
that~(\ref{4.7}) is equivalent to
	\begin{equation}	\label{5.01}
\mmor{U}_\gamma(t,s) =
\boldsymbol{l}_{\gamma(t)}^{-1}(t)
\mope{U}(t,s)\boldsymbol{l}_{\gamma(s)}(s).
	\end{equation}

	Let
$\boldsymbol{\Omega}(x):=\left[\Omega_{a}^{{\ }b}(x)\right]$
and
$\boldsymbol{\omega}(t):=\left[\omega_{a}^{{\ }b}(t)\right]$
be nondegenerate matrices. The changes
\[
\{e_a(x)\} \to \{ e_a^\prime (x):=\Omega_{a}^{{\ }b}(x)e_b(x) \} %\]
%and \[
,\qquad
\{f_a(t)\} \to \{ e_a^\prime (t):=\omega_{a}^{{\ }b}(t)e_b(t) \}
\]
of the bases in $\bHil_x$ and $\Hil$, respectively, lead to the
transformation of the matrices of the components of $\Phi_x\in \bHil_x$ and
$\phi\in\Hil$, respectively, according to
\[
\boldsymbol{\Phi}_x\mapsto \boldsymbol{\Phi}_{x}^{\prime} =
\left(\boldsymbol{\Omega}^\top(x)\right)^{-1}\boldsymbol{\Phi}_x,\quad
\boldsymbol{\phi}\mapsto \boldsymbol{\phi}^{\prime} =
\left(\boldsymbol{\omega}^\top(t)\right)^{-1}\boldsymbol{\phi}.
\]
Here the super script $\top$ means matrix transposition, for example
\(
\boldsymbol{\Omega}^\top(x) :=
\left[ \left({\Omega}^\top(x)\right)_{{\ }b}^{a} \right]
\)
with
\(
\left({\Omega}^\top(x)\right)_{{\ }b}^{a} :=
{\Omega}_{b}^{{\ }a} (x) .
\)
  One easily verifies the transformation
	\begin{equation}	\label{5.02}
\boldsymbol{l}_x(t)\mapsto  \boldsymbol{l}_{x}^{\prime}(t) =
\left( \boldsymbol{\omega}^\top(t) \right)^{-1}
\boldsymbol{l}_x(t)
\boldsymbol{\Omega}^\top(x)
	\end{equation}
of the components of the linear isomorphisms $l_x\colon \bHil_x\to\Hil$ under
the above changes of bases.

	For any operator $\ope{A}(t)\colon\Hil\to\Hil$ we have
	\begin{equation}	\label{5.02a}
\mope{A}(t)\mapsto\mope{A}^\prime(t)
	= \left(\boldsymbol{\omega}^\top(t)\right)^{-1}
	  \mope{A}(t)
	  \boldsymbol{\omega}^\top(t).
	\end{equation}

	Analogously, if  $\mor{A}(t)$  is a morphism of $\HilB$, i.e.\ if
$\mor{A}\colon\bHil\to\bHil$ and $\pi\circ\mor{A}=\id_\base$, and
$\mor{A}_x:=\left.\mor{A}(t)\right|_{\bHil_x}$, then
	\begin{equation}	\label{5.02b}
\mmor{A}_x(t)\mapsto\mmor{A}^\prime_x(t)
	= \left(\boldsymbol{\Omega}^\top(t)\right)^{-1}
	  \mmor{A}_x(t)
	  \boldsymbol{\Omega}^\top(t).
	\end{equation}

	Note that the components of $\ope{U}(t,s)$, when referred to a pair
of bases $\{e_a(t)\}$ and $\{e_a(s)\}$, transform according to
	\begin{equation}	\label{5.03}
\mope{U}(t,s) \mapsto \mope{U}^\prime(t,s) =
\left( \boldsymbol{\omega}^\top(t) \right)^{-1}
\mope{U}(t,s)
\boldsymbol{\omega}^\top(s).
	\end{equation}
Analogously, the change
\( \{e_a(\gamma(t))\} \to \{ e_a^\prime (t;\gamma) :=
\Omega_{a}^{{\ }b}(t;\gamma)e_b(\gamma(t)) \} \),
with a nondegenerate matrix
$\boldsymbol{\Omega}(t;\gamma):=\left[ \Omega_{a}^{{\ }b}(t;\gamma) \right]$
along $\gamma$, implies
\footnote{%
Cf.~\cite[equation~\eref{+:2.10}]{bp-normalF-LTP}
or~\cite[equation~(4.10)]{bp-LTP-general}, where the notation
$ \Mat{L}(t,s;\gamma)=H(t,s;\gamma)=\mmor{U}_\gamma(s,t;\gamma) $
and
$ \mor{A}(t)=\boldsymbol{\Omega}^\top(t;\gamma)$
is used.%
}
	\begin{equation}	\label{5.04}
\mmor{U}_\gamma(t,s) \mapsto \mmor{U}_\gamma^\prime(t,s) =
\left( \boldsymbol{\Omega}^\top(t;\gamma)\right)^{-1}
\mmor{U}_\gamma(t,s)
\boldsymbol{\Omega}^\top(s;\gamma).
	\end{equation}

	Substituting $\psi(t)=\psi^a(t)f_a(t)$ into~(\ref{2.5}), we get the
\emph{matrix Schr\"odinger equation}
	\begin{equation}	\label{5.05}
\frac{d\boldsymbol{\psi}(t)}{dt} =
\mHam(t) \boldsymbol{\psi}(t)
	\end{equation}
where
	\begin{equation}	\label{5.06}
\mHam(t) :=
\HamM(t) - \ih \boldsymbol{E}(t)
	\end{equation}
is the \emph{matrix Hamiltonian} (in the Hilbert space description).
	Here $\boldsymbol{E}(t)=\left[E_{a}^{{\ }b}(t)\right]$
determines the expansion of
%
% this is well defined as $f_a$ are in $\Hil$ and not in $\bHil$!!!!
%
${d f_a(t)}/{dt}$ over $\{f_a(t)\}\subset\Hil$, that is
${d f_a(t)}/{dt}=E_{a}^{{\ }b}(t) f_b(t)$; if $f_a(t)$ are independent
of $t$, which is the usual case,
we have $\boldsymbol{E}(t)=0$. In the last case
$\mHam = \HamM$.
It is important to be noted that $\mHam$ is
independent of
$\boldsymbol{E}(t)$. In fact, applying~(\ref{2.7}) to the basic vector
$f_a(t)$, we get
\(
\Ham(t) f_a(t) =
\ih [ (\frac{\partial}{\partial t} \ope{U}(t,t_0) ) f_b(t_0) ]
\ope{U}_{a}^{{\ }b}(t_0,t) =
\ih [ \frac{\partial}{\partial t} ( f_c(t) \ope{U}_{b}^{{\ }c}(t,t_0) ) ]
\ope{U}_{a}^{{\ }b}(t_0,t)
\),
that is
	\begin{equation}	\label{5.07}
\HamM(t) =
\ih \frac{\partial \mope{U}(t,t_0)}{\partial t}
\mope{U}(t_0,t) + \ih \boldsymbol{E}(t)
	\end{equation}
which leads to
	\begin{equation}	\label{5.08}
\mHam(t) =
\ih \frac{\partial \mope{U}(t,t_0)}{\partial t}
\mope{U}(t_0,t).
	\end{equation}

	Substituting into~(\ref{5.05}) the matrix form of~(\ref{4.3}), we
find the \emph{matrix-bundle Schr\"odinger equation}

	\begin{equation}	\label{5.1}
\ih\frac{d \boldsymbol{\Psi}_\gamma(t) }{dt} =
\mbHam_{\gamma}(t) \boldsymbol{\Psi}_\gamma(t)
	\end{equation}
where the
\emph{matrix-bundle Hamiltonian}
is
	\begin{equation}	\label{5.2}
\mbHam_{\gamma}(t)  :=
\boldsymbol{l}_{\gamma(t)}^{-1}(t)
\HamM(t) \boldsymbol{l}_{\gamma(t)}(t)
 -
\ih \boldsymbol{l}_{\gamma(t)}^{-1}(t)
\left(
\frac{d \boldsymbol{l}_{\gamma(t)}(t) }{dt} +
\boldsymbol{E}(t) \boldsymbol{l}_{\gamma(t)}(t)
\right).
	\end{equation}

	Combining~(\ref{5.06}) and~(\ref{5.2}), we find the following
connection between the conventional and bundle matrix Hamiltonians:
	\begin{equation}	\label{5.2'}
\mbHam_{\gamma}(t)  =
\boldsymbol{l}_{\gamma(t)}^{-1}(t)
\HamM^{\mathbf{m}}(t) \boldsymbol{l}_{\gamma(t)}(t)
-
\ih \boldsymbol{l}_{\gamma(t)}^{-1}(t)
\frac{d \boldsymbol{l}_{\gamma(t)}(t)}{dt}.
	\end{equation}

	\begin{rem}{5.1}
	Choosing $e_a(x)=l_x^{-1}(f_a)$ for
$df_a(t)/dt=0$, we get
$\boldsymbol{l}_x(t)=\left[ \delta_{a}^{b} \right]$. Then
$\bHamM_{\gamma}(t) = \HamM(t)$. So, as
$\Ham^\dag=\Ham$, we have
\(
\left(\mbHam_{\gamma}(t)\right)^\dag =
\HamM^\dag(t) =
\HamM(t) = \mbHam_{\gamma}(t)
\)
where we use the dagger ($\dag$) to denote also matrix Hermitian
conjugation.
	Here $\mbHam_{\gamma}(t)$
is a Hermitian matrix in the chosen basis,
but in other bases it may not be such (see~(\ref{5.12}) below).
Analogously, choosing $\{f_a(t)\}$ such that $\boldsymbol{E}(t)=0$,
we see that
$\mHam(t)=\HamM(t)$
is a Hermitian matrix, otherwise it may not be such.
	\end{rem}

	\begin{rem}{5.2}
	Note that, due to~(\ref{5.2'}), the transition
$\mHam\to\mbHam_{\gamma}$
is very much alike a gauge (or connection)
transformation~\cite{Konopleva-Popov}
(see also below~(\ref{5.10})--(\ref{5.12})).
	\end{rem}

	Because of~(\ref{5.1}) and~(\ref{4.4}) there is 1:1 correspondence
between $\mmor{U}_\gamma$ and $\mbHam_{\gamma}$
expressed through the initial-value problem (cf.~(\ref{2.6}))
	\begin{equation}	\label{5.3}
\ih \frac{\partial\mmor{U}_\gamma(t,t_0)}{\partial t} =
\mbHam_{\gamma}(t) \mmor{U}_\gamma(t,t_0), \qquad
\mmor{U}_\gamma(t_0,t_0) = \openone_{\bHil_{\gamma(t_0)}},
	\end{equation}
or via the equivalent to it integral equation
	\begin{equation}	\label{5.4}
\mmor{U}_\gamma(t,t_0) = \openone_{\bHil_{\gamma(t_0)}} +
\iih
\int\limits_{t_0}^{t}
\mbHam_{\gamma}(\tau)
\mmor{U}_\gamma(\tau,t_0)
d\tau.
	\end{equation}

	So, if $\mbHam_{\gamma}$ is given, we have
(cf.~(\ref{2.8}))
	\begin{equation}	\label{5.5}
\mmor{U}_\gamma(t,t_0) =
\Texp
\int\limits_{t_0}^{t}
\iih
\mbHam_{\gamma}(\tau)
d\tau
	\end{equation}
and, conversely, if $\mmor{U}_\gamma$ is given, then (cf.~(\ref{2.7})
and~(\ref{5.08}))
\footnote{%
Expressions like
$ ({\partial\ope{U}(t,t_0)}/{\partial t} ) \ope{U}(t_0,t)$,
$( {\partial\mmor{U}_\gamma(t,t_0)}/{\partial t} )
\mmor{U}_\gamma^{-1}(t,t_0)$,
and
$\ope{U}{(t,t_0)}\ope{U}(t_0,t_1)$
are independent of $t_0$
due to~\cite[propositions~\ref{+:Prop2.1}
		and~\ref{+:Prop2.4}]{bp-normalF-LTP}
or~\cite[propositions~2.1 and~2.4]{bp-LTP-general}
(see also~(\ref{3.4}) and~\cite[lemma~3.1]{bp-TP-general}).%
}
	\begin{equation}	\label{5.6}
\mbHam_{\gamma}(t) =
\ih
\frac{\partial\mmor{U}_\gamma(t,t_0)}{\partial t}
\mmor{U}_\gamma^{-1}(t,t_0) =
\frac{\partial\mmor{U}_\gamma(t,t_0)}{\partial t}
\mmor{U}_\gamma^{}(t_0,t).
	\end{equation}

	The next step is to write the above matrix equations into an
invariant, i.e.\  basis-independent, form. For this purpose we shall use the
introduce in~\cite{bp-normalF-LTP,bp-LTP-general} derivation along paths
uniquely corresponding to any linear transport along paths in a vector
bundle.

	Let $\mor{D}$ be the derivation along paths corresponding to the
bundle evolution transport $\mor{U}$, that is
(cf.~\cite[definition~\ref{+:Defn2.3}]{bp-normalF-LTP}
or~\cite[definition~4.1]{bp-LTP-general})
 $\mor{D}\colon \gamma\mapsto \mor{D}^\gamma$,
where $\mor{D}^\gamma$, called derivation along $\gamma$ generated by
$\mor{U}$, is such that
$\mor{D}^\gamma\colon s \mapsto \mor{D}^\gamma_s$
and the derivation
\[
\mor{D}^\gamma_s
 \colon
\Sec^1\left(\left.\HilB\right|_{\gamma(J)}\right)
  \to
\proj^{-1}(\gamma(s))
\]
along $\gamma$ at $s$ generated by $\mor{U}$ is defined by
	\begin{equation}	\label{5.7}
% \left(\mor{D}^\gamma(\chi)\right)(\gamma(s)):=
\mor{D}_{s}^{\gamma}\chi
  := \lim_{\varepsilon\to0}
	\left\{
\frac{1}{\varepsilon}
	\left[
\mor{U}_\gamma(s,s+\varepsilon)\chi(\gamma(s+\varepsilon)) - \chi(\gamma(s))
	\right]
	\right\}
	\end{equation}
for any $C^1$ section $\chi$ over $\gamma(J)$ in $\HilB$.

	By~\cite[equation~\eref{+:2.22}]{bp-normalF-LTP}
or~\cite[proposition~4.2]{bp-LTP-general}
the local explicit form of~(\ref{5.7}) is
	\begin{equation}	\label{5.8}
\mor{D}_{s}^{\gamma}\chi =
\left(
\frac{d\chi^a(\gamma(s))}{ds} +
\Gamma_{{\ }b}^{a}(s;\gamma)\chi^b(\gamma(s))
\right)
e_a(\gamma(s))
	\end{equation}
where the
\emph{coefficients}
$\Gamma_{{\ }a}^{b}(s;\gamma)$ of $\mor{U}$ are defined by
	\begin{equation}	\label{5.9}
\Gamma_{{\ }a}^{b}(s;\gamma) :=
\left.
\frac{\partial\left(\mor{U}_\gamma(s,t)\right)_{{\ }a}^{b}}{\partial t}
\right|_{t=s} =
- \left.
\frac{\partial\left(\mor{U}_\gamma(t,s)\right)_{{\ }a}^{b}}{\partial t}
\right|_{t=s} .
	\end{equation}

	Using~(\ref{4.6}) and~(\ref{5.6}), both for $t_0=t$, we see that
	\begin{equation}	\label{5.10}
\boldsymbol{\Gamma}_\gamma(t) :=
\left[ \Gamma_{{\ }a}^{b}(t;\gamma) \right] =
- \iih \mbHam_{\gamma}(t)
	\end{equation}
which expresses a fundamental result:
\emph{up to a constant the matrix-bundle Hamiltonian coincides with the
matrix of coefficients of the bundle evolution transport}
(in a given field of bases). Let us recall that, using another arguments,
analogous result was obtained in~\cite[sect.~5]{bp-BQM-preliminary}.

	There are two invariant operators corresponding to the Hamiltonian
$\Ham$ in $\Hil$: the bundle-evolution transport $\mor{U}$ and the
corresponding to
it derivation along paths $\mor{D}$. The equations~(\ref{5.1})--(\ref{5.10}), as
well as the general results of~\cite[\S~\ref{+:Sect2}]{bp-normalF-LTP}
and~\cite[\S~4]{bp-LTP-general}, imply that these
three operators, namely $\Ham$, $\mor{U}$, and $\mor{D}$, are equivalent in a
sense that  if one of them is given, then the remaining ones are uniquely
determined.

	\begin{exmp}{5.0}
Let $\{e_a(x)\}$  be fixed by $e_a(x)=l_{x}^{-1}(f_a)$ for $df(t)/dt=0$.
Then
$\mbHam_{\gamma}(t)$ is a Hermitian matrix (see
remark~\ref{rem5.1}). Consequently, in this case,
$\boldsymbol{\Gamma}_\gamma(t)$ is anti-Hermitian, i.e.\
\(
\left(\boldsymbol{\Gamma}_\gamma(t)\right)^\dag =
- \boldsymbol{\Gamma}_\gamma(t).
\)
Note that for other choices of the bases this property may not hold.
	\end{exmp}

	\begin{exmp}{5.1}
	Let $\Ham$ be given and independent of $t$, i.e.\
$\partial \Ham(t)/\partial t =0$, and $\{e_a(x)\}$  be fixed by
$e_a(x)=l_{x}^{-1}(f_a)$ for $df(t)/dt=0$. Then
$\boldsymbol{l}_x(t)=\left[\delta_{a}^{b}\right]$ with
$\delta_{a}^{b}$ defined above.
Equations~(\ref{5.2}) and~(\ref{5.10}) yield
$\mbHam_{\gamma}(t)=\HamM(t)$ and
$\boldsymbol{\Gamma}_\gamma(t)=-\HamM(t)/\ih$.
Finally, now the solution of~(\ref{5.3}) is
\(
\mmor{U}_\gamma(t,t_0) =
\exp \left( \HamM(t)(t-t_0)/\ih \right)
\)
(cf.~(\ref{5.5})).
	\end{exmp}

	According to~\cite[equation~\eref{+:2.26}]{bp-normalF-LTP}
(or~\cite[equation~(4.11)]{bp-LTP-general}) and
footnote~\ref{L-transport:evolution-transport}%
% on page~\pageref{L-transport:evolution-transport}%
, if a basis
$\{e_a(\gamma(t))\}$ is change to
$\{ e_{a}^{\prime}(t;\gamma) =
\Omega_{a}^{{\ }b}(t;\gamma) e_b(\gamma(t)) \}$
with $\det\boldsymbol{\Omega}(t;\gamma)\not=0$,
\(
\boldsymbol{\Omega}(t;\gamma) :=
\left[\Omega_{a}^{{\ }b}(t;\gamma)\right],
\)
then $\boldsymbol{\Gamma}_\gamma(t)$ transforms into%
\footnote{%
In~\cite{bp-normalF-LTP,bp-LTP-general} the matrix
$\mor{A}(t)=\boldsymbol{\Omega}^\top(t;\gamma)$ is used
instead of $\boldsymbol{\Omega}(t;\gamma)$.%
}
	\begin{equation}	\label{5.11}
\boldsymbol{\Gamma}_\gamma^\prime(t) =
(\boldsymbol{\Omega}^{\top}(t;\gamma))^{-1}
\boldsymbol{\Gamma}_\gamma(t)
\boldsymbol{\Omega}^\top(t;\gamma) +
(\boldsymbol{\Omega}^{\top}(t;\gamma))^{-1}
\frac{d \boldsymbol{\Omega}^\top(t;\gamma)}{dt}.
	\end{equation}
This result is also a corollary of~(\ref{5.03}) and~(\ref{5.9}).

	Hence (see~(\ref{5.10})), the matrix-bundle Hamiltonian undergoes
the change
$\mbHam_{\gamma}(t) \mapsto \,^{\prime}\!\mbHam_{\gamma}(t)$
where
	\begin{equation}	\label{5.12}
\,^{\prime}\!\mbHam_{\gamma}(t) =
(\boldsymbol{\Omega}^\top(t;\gamma))^{-1}
\mbHam_{\gamma}(t)
\boldsymbol{\Omega}^\top(t;\gamma)
-\ih
(\boldsymbol{\Omega}^\top(t;\gamma))^{-1}
\frac{d \boldsymbol{\Omega}^\top(t;\gamma)}{dt},
	\end{equation}
which can be deduced from~(\ref{5.2'}) too.

	Now we are able to write into an invariant form the matrix-bundle
Schr\"odinger equation~(\ref{5.1}). Substituting~(\ref{5.10}) into~(\ref{5.1})
and using~(\ref{5.8}), we find that~(\ref{5.1}) is equivalent to
	\begin{equation}	\label{5.13}
\mor{D}_{t}^{\gamma}\Psi_\gamma = 0.
	\end{equation}
	This is the (invariant)
\emph{bundle Schr\"odinger equation}
(for the state sections).
Since it coincides with the
\emph{linear transport equation along
$\gamma$}~\cite[definition~5.2]{bp-LTP-appl}
for the bundle evolution transport,
it has a very simple and fundamental geometrical meaning.
By~\cite[proposition~5.4]{bp-LTP-appl} this
is equivalent to the statement that
\emph{$\Psi_\gamma$ is a (linearly) transported along $\gamma$ section}
with respect to the bundle evolution transport (expressed in other terms
via~(\ref{4.4}); see~\cite[definition~2.2]{bp-TP-general}).
Note that~(\ref{5.13}) and~(\ref{4.4}) are compatible
as~\cite[equation~(4.5)]{bp-LTP-general} is fulfilled
(see also~\cite[equation~\eref{+:2.20}]{bp-normalF-LTP}):
\(
\mor{D}_{t}^{\gamma}\circ \mor{U}_\gamma(t,t_0)\equiv0,\ t,t_0\in J
\)
($\gamma$ is not a summation index here!).
	Moreover, if $\mor{D}$ is given (independently of $\mor{U}$, e.g.
through~(\ref{5.8})), then
from~\cite[proposition~5.4]{bp-LTP-appl} follows that $\mor{U}$ is the unique
solution of the (invariant) initial-value problem
	\begin{equation}	\label{5.14}
\mor{D}_{t}^{\gamma}\circ \mor{U}_\gamma(t,t_0) = 0
\qquad \mor{U}_\gamma(t_0,t_0) = \id_{\bHil_{\gamma(t_0)}}.
	\end{equation}
This is the bundle Schr\"odinger equation for the evolution transport
(operator) $\mor{U}$. In fact it is the inversion of~(\ref{5.7}) with
respect to $\mor{U}$.

	Let us summarize in conclusion. There are two equivalent ways of
describing the unitary evolution of a quantum system:
(i) through the evolution transport $\ope{U}$ (see~(\ref{2.1})) or by the
Hermitian Hamiltonian $\Ham$ (see~(\ref{2.5})) in the Hilbert space $\Hil$
(which is the typical fibre in the bundle description) and
(ii) through the bundle evolution transport $\mor{U}$ (see~(\ref{4.4})),
which is a Hermitian (and unitary) transport along paths, or the derivation
along paths $\mor{D}$ (see~(\ref{5.13})) in the Hilbert fibre bundle $\HilB$.
In the bundle description $\mor{U}$ corresponds to $\ope{U}$
(see~(\ref{4.7})) and $\mor{D}$ to $\Ham$ (see~(\ref{5.8}) and~(\ref{5.10})).

	Since now we have in our disposal the machinery required for analysis
of~\cite{Asorey&Carinena&Paramio}, we, as promised in
Sect.~\ref{introduction-I}, want to make some comments on it.
In~\cite[p.~1455, left column, paragraph~4]{Asorey&Carinena&Paramio} is
stated ``that in the Heisenberg gauge (picture) the Hamiltonian is the null
operator''. If so, all eigenvalues of the Hamiltonian vanish and, as they are
picture\ndash independent, they are null in any picture of quantum mechanics.
Consequently, form here one deduces the absurd conclusion that the `energy
levels of any system coincide and correspond to one and the same energy equal
to zero'. Since the paper~\cite{Asorey&Carinena&Paramio} is mathematically
completely correct and rigorous, there is something wrong with the physical
interpretation of the mathematical scheme developed in it. Without
going into details, we describe below the solution of this puzzle which
simultaneously throws a bridge between~\cite{Asorey&Carinena&Paramio} and
the present investigation.

	In~\cite{Asorey&Carinena&Paramio} the system's Hilbert space
$\mathfrak{H}$ is replace by a differentiable Hilbert bundle
$\mathfrak{E}(\mathbb{R}_+,\mathfrak{H})$
(in our terms $(\mathfrak{E},\pi,\mathbb{R}_+)$ with a fibre $\mathfrak{H})$),
$\mathbb{R}_+:=\{t :  t\in\mathbb{R}, t\ge0\}$,
which is an associated Hilbert bundle of the principle fibre bundle
$\mathfrak{P}\bigl(\mathbb{R}_+,\mathfrak{U}(\mathfrak{H})\bigr)$ of orthonormal bases of
$\mathfrak{H}$ where $\mathfrak{U}(\mathfrak{H})$ is the unitary group of (linear)
bounded invertible operators in $\mathfrak{H}$ with bounded inverse.
Let $p:\mathfrak{U}(\mathfrak{H})\to GL(\mathbb{C}, dim\mathfrak{H})$ be a (linear and
continuous) representation of $\mathfrak{U}(\mathfrak{H})$ into the general
linear group of $dim\mathfrak{H}$\ndash dimensional matrices. An obvious
observation is that~\cite[equation~(4.6)]{Asorey&Carinena&Paramio} under $p$
transforms, up to notation, to our equation~\eref{5.12}
(in~\cite{Asorey&Carinena&Paramio} is taken $\hbar=1$). Thus we see that
what in~\cite{Asorey&Carinena&Paramio} is called Hamiltonian is actually
the (analogue of the) matrix\ndash bundle Hamiltonian $\mbHam_\gamma(t)$,
not the Hamiltonian $\Ham$ itself (see also Sect.~\ref{VI}). This
immediately removes the above\ndash pointed conflict: as we shall see later
in the third part of this series, along any $\gamma$ (or,
over $\mathbb{R}_+$ in the notation of~\cite{Asorey&Carinena&Paramio} - see
below), we can choose a field of frames (bases) in which $\mbHam_\gamma(t)$
identically vanishes but, due to~\eref{5.2}, this does not imply the
vanishment of the Hamiltonian at all. This particular choice of the frames
along $\gamma$  corresponds to the `Heisenberg gauge'
in~\cite{Asorey&Carinena&Paramio}, normally known as Heisenberg picture.

	Having in mind the above, we can
describe~\cite{Asorey&Carinena&Paramio} as follows. In it we have
$\bHil=\mathfrak{E}$, $\base=\mathbb{R}_+$, $\Hil=\mathfrak{H}$ (the conventional
system's Hilbert space), $J=\mathbb{R}_+$, $\gamma=\id_{\mathbb{R}_+}$
(other choices of $\gamma$  correspond to reparametrization of the time), and
$\frac{\pd}{\pd t},\ t\in\mathbb{R}_+$ is the analog of $D^+$
in~\cite{Asorey&Carinena&Paramio}. As we already pointed, the
matrix\ndash bundle Hamiltonian $\mbHam_\gamma(t)$ represents the operator
$A(t)$ of~\cite{Asorey&Carinena&Paramio}, incorrectly identified there with
the `Hamiltonian' and the choice of a field of bases over
$\gamma(J)=\mathbb{R}_+=\base$ corresponds to an appropriate `choice of the
gauge' in~\cite{Asorey&Carinena&Paramio}. Now, after its correspondence
between~\cite{Asorey&Carinena&Paramio} and the present work is set, one can
see that under the representation $p$ the main results
of~\cite{Asorey&Carinena&Paramio}, expressed
by~\cite[equations~(4.5), (4.6) and~(4.8]{Asorey&Carinena&Paramio},
correspond to our equations~\eref{5.13} (see also~\eref{5.8}), \eref{5.12}
and~\eref{5.02b} respectively.

	Ending with the comment on~\cite{Asorey&Carinena&Paramio}, we note
two things. First, this paper uses a rigorous mathematical base, analogous to
the one in~\cite{Prugovecki-QMinHS}, which is not a goal of our work. And,
second, the ideas of~\cite{Asorey&Carinena&Paramio} are a very good motivation
for the present investigation and are helpful for its better understanding.

\section {The bundle description of observables}
\label{VI}
\setcounter{equation} {0}

	In quantum mechanics is accepted that to any dynamical variable,
say $\dyn{A}$, there corresponds a unique observable, say
$\ope{A}(t)$, which is a Hermitian linear operator in the Hilbert space
$\Hil$, i.e.\  $\ope{A}(t)\colon \Hil\to\Hil$, $\ope{A}(t)$ is linear, and
$\ope{A}^\dag=\ope{A}$~\cite{Messiah-QM,Prugovecki-QMinHS,Fock-FQM}.
What is the analogue of $\ope{A}(t)$ in the developed here bundle description?
Below we prove that this is a suitable bundle morphism
\(\mor{A}\)
of the introduced in Sect.~\ref{IV} fibre bundle $\HilB$.

	Let $\psi^{(\lambda)}(t)\in\Hil$ be an eigenvector of $\ope{A}(t)$
with eigenvalue $\lambda$ ($\in\mathbb{R}$), i.e.\
$\ope{A}(t)\psi^{(\lambda)}(t) = \lambda \psi^{(\lambda)}(t)$.
According to~(\ref{4.3}) to $\psi^{(\lambda)}(t)$ corresponds the vector
\(
\Psi_\gamma^{(\lambda)}(t) =
l_{\gamma(t)}^{-1} \psi^{(\lambda)}(t) \in \bHil_{\gamma(t)}
\)
in the bundle description. But the Hilbert space and Hilbert bundle
descriptions of a quantum evolution are fully equivalent
(see Sect.~\ref{IV}). Hence to $\ope{A}(t)$
in $\bHil_{\gamma(t)}$ must correspond certain operator which we denote by
$\mor{A}_{\gamma}(t)$. We define this operator by
demanding any
$\Psi_\gamma^{(\lambda)}(t)$
to be its eigenvector with eigenvalue $\lambda$, i.e.\
\(
\left( \mor{A}_{\gamma}(t) \right)
				\Psi_\gamma^{(\lambda)}(t) :=
\lambda \Psi_\gamma^{(\lambda)}(t).
\)
Combining this equality with the preceding two, we easily verify that
\(
\bigl( \mor{A}_{\gamma}(t) \circ
			l_{\gamma(t)}^{-1} \bigr)
\psi^{(\lambda)}(t) =
\bigl(l_{\gamma(t)}^{-1} \circ \ope{A}(t) \bigr) \psi^{(\lambda)}(t)
\)
where the linearity of $l_x$ has been used. Admitting that
$\{\psi^{(\lambda)}(t)\}$ form a complete set of vectors, i.e a basis of
$\Hil$, we find
	\begin{equation}	\label{6.0}
\mor{A}_{\gamma}(t) =
l_{\gamma(t)}^{-1} \circ \ope{A}(t) \circ l_{\gamma(t)}
\colon  \bHil_{\gamma(t)}\to\bHil_{\gamma(t)}.
	\end{equation}

	More `physically' the same result is derivable from~(\ref{2.9}) too.
The mean value
\( \langle \ope{A} \rangle_{\psi}^{t} \) of \(\ope{A}\)
	at a state \(\psi(t)\) is given
by~(\ref{2.9}) and the mean value of
\( \mor{A}_{\gamma}(t) \) at a state \(\Psi_\gamma(t) \)
is
	\begin{equation}	\label{6.1'}
\left\langle
\mor{A}_{\gamma}(t)
\right\rangle_{\Psi_\gamma}^{t} =
\frac{
\langle \Psi_\gamma(t) | \mor{A}_\gamma(t) \Psi_\gamma(t) \rangle_{\gamma(t)}
}
{
\langle \Psi_\gamma(t) | \Psi_\gamma(t) \rangle_{\gamma(t)}
},
	\end{equation}
that is it is given via~(\ref{2.9}) in which the scalar
product $\langle\cdot | \cdot\rangle_x$, defined by~(\ref{4.8}), is used
instead of
$\langle\cdot | \cdot\rangle$.
	We define
$\mor{A}_{\gamma}(t)$ by demanding
	\begin{equation}	\label{6.1''}
\langle \ope{A}(t)\rangle_{\psi}^{t} =
\langle \mor{A}_{\gamma}(t) \rangle_{\Psi_\gamma}^{t}.
	\end{equation}
Physically this condition is very natural as it means that the observed
values of the dynamical variables are independent of the way we
calculate them. From this equality, (\ref{4.3}), and~(\ref{4.8}), we get
\(
\langle \psi(t) | \ope{A}(t)\psi(t) \rangle =
\langle \psi(t) |
l_{\gamma(t)}^{}
\circ \mor{A}_{\gamma}(t) \circ
l_{\gamma(t)}^{-1} \psi(t)
\rangle
\)
which, again, leads to~(\ref{6.0}). Thus we have also proved the equivalence
of~(\ref{6.0}) and~(\ref{6.1''}).

	According to equation~(\ref{6.0}), along $\gamma\colon J\to\base$
to any operator $\ope{A}(t)\colon \Hil\to\Hil$,  $t\in J$, there
corresponds a unique map
$\mor{A}_\gamma(t)\colon  \bHil_{\gamma(t)}\to\bHil_{\gamma(t)}$
in any fibre $\bHil_{\gamma(t)}$, $t\in J$, in $\HilB$. If
$J^\prime\subseteq J$ is a subinterval on which  $\gamma$ is without
self-intersections and
\(
\mor{A}_{\gamma|J^\prime}\colon
\proj^{-1}(\gamma(J^\prime)) \to \proj^{-1}(\gamma(J^\prime))
\)
is defined by
\(
\left.\mor{A}_{\gamma|J^\prime}\right|_{\bHil_{\gamma(t^\prime)}}
=\mor{A}_\gamma(t^\prime)
\colon  \bHil_{\gamma(t^\prime)}\to\bHil_{\gamma(t^\prime)},
\)
 $t^\prime\in J^\prime$,
then
$\mor{A}_{\gamma|J^\prime}\in\left(\Morf\HilB|_{\gamma(J^\prime)}\right)$,
i.e.\  $\mor{A}_{\gamma|J^\prime}$ is a morphism on the restricted on
$\gamma(J^\prime)$ fibre bundle $\HilB$. In the general case we define the
multiple-valued map $\mor{A}_\gamma\colon \bHil\to\bHil$ via
\(
\left.\mor{A}_\gamma\right|_{\bHil_x} :=
\{ \mor{A}_\gamma(t) : t\in J,\quad \gamma(t)=x \}
\)
for every  $x\in\base$.
Evidently
$\left.\mor{A}_\gamma\right|_{\bHil_x} = \varnothing$
for $x\not\in\gamma(J)$ and
\(
\left.\mor{A}_\gamma(t)\right|_{\bHil_{\gamma(t)}} \colon
\bHil_{\gamma(t)}\to\bHil_{\gamma(t)},\ t\in J,
\)
the multiplicity of
$\left.\mor{A}_\gamma\right|_{\bHil_{\gamma(t)}} $
being equal to one plus the number of self-intersections of $\gamma$  at the
point $\gamma(t)$.
We call a \emph{(bundle) morphism along paths}%
\footnote{%
Cf. the definition of a (bundle) morphism
\(
\mor{C}\in\Morf\HilB:=\{\mor{B}:\quad \mor{B}\colon \bHil\to \bHil,\
\proj\circ \mor{B}=\id_{\base} \}
\)
of $\HilB$.%
}
any map $\mor{A}\colon \gamma\mapsto\mor{A}_\gamma$, where
$\mor{A}_\gamma\colon \bHil\to\bHil$ can be multiple-valued and such that
\(
\proj\circ\left(
\left.\mor{A}_\gamma\right|_{\proj^{-1}(\gamma(J))}
\right)
=\id_{\gamma(J)}
\)
and $\left.\mor{A}_\gamma\right|_{\bHil_x}=\emptyset$
for $x\not\in\gamma(J)$.
We call the (possibly multiple-valued) map $\mor{A}_\gamma$  a
\emph{(bundle) morphism along the path $\gamma$}.
Hence, the above-defined map $\mor{A}_\gamma$ is a morphism along $\gamma$
which is singled-valued (and consequently a morphism over $\gamma(J)$) iff
$\gamma$ is without self-intersections. Therefore the map
$\mor{A}\colon \gamma\mapsto\mor{A}_\gamma$ is morphism along paths. We call
$\mor{A}$ \emph{Hermitian} and denote this by $\mor{A}^\ddagger=\mor{A}$, if
$\mor{A}_\gamma$ are such, i.e.\  if~(\ref{4.morphism}) holds for
$\mor{A}_\gamma$ instead of $\mor{A}$. The morphism along paths $\mor{A}$ is
Hermitian if $\ope{A}(t)$ is a Hermitian operator, viz.\ we have
	\begin{equation}	\label{6.2}
\mor{A}^\ddag = \mor{A}   \iff
\mor{A}_\gamma^\ddag(t) = \mor{A}_\gamma(t)   \iff
\ope{A}^\dag(t) = \ope{A}(t) ,
	\end{equation}
which is a simple corollary of~(\ref{6.0}) and~(\ref{4.15}).
Hence, if the morphism $\mor{A}_\gamma(t)$ along $\gamma$ corresponds to an
observable $\ope{A}$, it is Hermitian because   $\ope{A}(t)$ is such by
assumption~\cite{Messiah-QM,Dirac-PQM}

	Consequently,
\emph{%
to any observable $\ope{A}$ there corresponds a unique
Hermitian bundle morphism $\mor{A}$ along paths and vice versa.%
}
Explicitly this correspondence is given by~(\ref{6.0}) which will be
assumed hereafter. Its consequence is the
\emph{independence of the
physically measurable quantities} (and the eigenvalues of the corresponding
operators) \emph{of the mathematical way we describe them}, as it should be.

	Generally to any operator $\ope{A}\colon \Hil\to\Hil$ there
corresponds a unique (global) morphism
$\overline{\mor{A}}\in\Morf\HilB$ given by
	\begin{equation}	\label{6.1}
\overline{\mor{A}}_x = \left.\overline{\mor{A}}\right|_{\bHil_{x}} =
l_{x}^{-1} \circ \ope{A} \circ l_{x}, \qquad x\in \base,
\quad \ope{A}\colon \Hil\to\Hil.
	\end{equation}
Consequently to an observable $\ope{A}(t)$  can be assigned the (global)
morphism $\overline{\mor{A}}(t)$,
 $\overline{\mor{A}}(t)|_{\bHil_x}=l_{x}^{-1}\circ \ope{A}(t)\circ l_x$.
But this morphism $\overline{\mor{A}}(t)$ is almost useless for our goals as
it simply gives in any fibre $\bHil_x$ a linearly isomorphic image of the
initial observable $\ope{A}(t)$ (see Sect.~\ref{new-I}).

	Notice that $\mor{A}_\gamma(t)$ generally
depends explicitly on $t$ even if $\ope{A}$ does not.
In fact, from~(\ref{6.0}) we get
	\begin{equation}	\label{6.6}
\frac{\partial \mmor{A}_\gamma(t) } {\partial t} =
\left[
\boldsymbol{g}_\gamma(t), \mmor{A}_\gamma(t)
\right]_{\_}
+
\boldsymbol{l}_{\gamma(t)}^{-1}(t)
\frac{\partial \mope{A}(t) }{\partial t}
\boldsymbol{l}_{\gamma(t)}(t),
	\end{equation}
where $[\cdot,\cdot]_-$ denotes the commutator of corresponding quantities
and
	\begin{equation}	\label{6.7}
\boldsymbol{g}_{\gamma}(t) := -  \boldsymbol{l}_{\gamma(t)}^{-1}(t)
\frac{d \boldsymbol{l}_{\gamma(t)}(t) }{d t} .
	\end{equation}

	In particular, to the Hamiltonian $\Ham$ in $\Hil$ there
corresponds the \emph{bundle Hamiltonian}
(or the \emph{bundle-Hamiltonian morphism along paths}) given by
	\begin{equation}	\label{6.2'}
\bHam_\gamma(t) :=
l_{\gamma(t)}^{-1} \circ \Ham(t) \circ l_{\gamma(t)}.
	\end{equation}
It is a Hermitian bundle morphism along paths,
$\bHam_\gamma^\ddag=\bHam_\gamma$, as $\Ham$ is a Hermitian operator.

	From~(\ref{6.2'}), using~(\ref{2.7}) and~(\ref{4.7}), we find
	\begin{equation}	\label{6.2'a}
\bHam_\gamma(t) = \ih l_{\gamma(t)}^{-1} \circ
\frac{\partial \ope{U}(t,t_0)}{\partial t} \circ l_{\gamma(t_0)} \circ
\mor{U}_\gamma(t_0,t).
	\end{equation}
	From here we can get a relationship between the matrix-bundle
Hamiltonian and the bundle Hamiltonian. For this purpose we
write~(\ref{6.2'a}) in a matrix form and using~(\ref{5.6}) and
\( {d f_a(t)}/{dt}=E_{a}^{{\ }b}f_b(t)\),
we obtain:
	\begin{equation}	\label{6.2''}
\bHamM_{\gamma}(t)  = \mbHam_{\gamma}(t) +
\ih \boldsymbol{l}_{\gamma(t)}^{-1}(t)
\left( \frac{d \boldsymbol{l}_{\gamma(t)} (t) }{dt} +
\boldsymbol{E}(t)\boldsymbol{l}_{\gamma(t)}(t)
\right).
	\end{equation}
Substituting here~(\ref{5.2}), we get
	\begin{equation}	\label{6.10'}
\bHamM_\gamma(t) =
\boldsymbol{l}_{\gamma(t)}^{-1}(t)
\HamM(t)\boldsymbol{l}_{\gamma(t)}(t)
	\end{equation}
which is simply the matrix form of~(\ref{6.2'}). Combining~(\ref{6.2''})
with~(\ref{5.2'}), we find the following connection between the matrix of
the bundle Hamiltonian and the matrix Hamiltonian:
	\begin{equation}	\label{6.10''}
\bHamM_\gamma(t) =
\boldsymbol{l}_{\gamma(t)}^{-1}(t)
\mHam(t)\boldsymbol{l}_{\gamma(t)}(t) +
\ih \boldsymbol{l}_{\gamma(t)}^{-1}(t)
\boldsymbol{E}(t)\boldsymbol{l}_{\gamma(t)}(t) .
	\end{equation}

    	We notice that, due to~(\ref{6.1}) as well as to~(\ref{6.0}), to the
identity map of $\Hil$ there corresponds a morphism along paths equal to the
identity map of $\bHil$:
	\begin{equation}	\label{6.2'''}
\id_\Hil \longleftrightarrow \id_\bHil .
	\end{equation}

	The results expressed by~(\ref{6.0}) and~(\ref{6.1}) enable functions
of observables in $\Hil$ to be transferred into ones of morphisms along paths
or morphisms of $\HilB$, respectively. We will illustrate this
in the case of, e.g., two variables. Let
\(
\ope{G}\colon (\ope{A}(t),\ope{B}(t)) \mapsto
 \ope{G}(\ope{A}(t),\ope{B}(t))\colon  \Hil\to\Hil
\)
be a function of the observables
$\ope{A}(t),\ope{B}(t)\colon \Hil\to\Hil$.
It is natural to define the bundle analogue $\mor{G}$ of $\ope{G}$ by
\[
\mor{G}\colon(\mor{A},\mor{B})\mapsto\mor{G}(\mor{A},\mor{B})\colon
\gamma\mapsto\mor{G}_\gamma(\mor{A},\mor{B})\colon
\proj^{-1}(\gamma(J))\to\proj^{-1}(\gamma(J)),
\]
where
 $\left.\mor{G}_\gamma(\mor{A},\mor{B})\right|_{\bHil_x}=\varnothing$
for $x\not\in\gamma(J)$ and
	\begin{multline}	\label{6.071}
\left. G_\gamma(\mor{A},\mor{B}) \right|_{\bHil_{\gamma(t)}} :=
l_{\gamma(t)}^{-1}\circ\ope{G}(\ope{A}(t),\ope{B}(t))\circ l_{\gamma(t)} \\
=
l_{\gamma(t)}^{-1}\circ
\ope{G}(l_{\gamma(t)}\circ\mor{A}_\gamma(t)\circ l_{\gamma(t)}^{-1} ,
	    l_{\gamma(t)}\circ\mor{B}_\gamma(t)\circ l_{\gamma(t)}^{-1})
\circ l_{\gamma(t)}.
	\end{multline}
Thus $G(\mor{A},\mor{B})$ is a bundle morphism along paths.
This definition becomes evident in the cases when $\ope{G}$ is a
polynom or if it is expressible as a convergent power series; in both
cases the multiplication has to be understood as an operator composition.
If we are dealing with one of these cases, the definition~(\ref{6.071})
follows from the fact that for any morphisms
$\mor{A}_1,\ldots,\mor{A}_k$, $k\in\mathbb{N}$
along paths of $\HilB$ the equality
	\begin{equation}	\label{6.072}
\mor{A}_{1,\gamma}(t)\circ
\mor{A}_{2,\gamma}(t)\circ \cdots \circ \mor{A}_{k,\gamma}(t) =
l_{\gamma(t)}^{-1}\circ (\ope{A}_1(t)\circ\ope{A}_2(t)\circ\cdots
\circ\ope{A}_k(t)) \circ l_{\gamma(t)}
	\end{equation}
holds due to~(\ref{6.0}). In these cases $\mor{G}(\mor{A},\mor{B})$ depends
only on $\mor{A}$ and $\mor{B}$ and it is explicitly independent on the
isomorphisms $l_x$, $x\in\base$.

	The commutator of two operators is a an important operator
function in quantum mechanics. In the Hilbert space and bundle
descriptions it is defined by
$[\ope{A},\ope{B}]_{\_}:=\ope{A}\circ\ope{B}-\ope{B}\circ\ope{A}$
and
$[\mor{A},\mor{B}]_{\_}:=\mor{A}\circ \mor{B} -\mor{B}\circ \mor{A}$
respectively. The relation
	\begin{equation}	\label{6.073}
[\mor{A}_\gamma(t),\mor{B}_\gamma(t)]_{\_} =
l_{\gamma(t)}^{-1} \circ [\ope{A},\ope{B}]_{\_} \circ l_{\gamma(t)}
	\end{equation}
is an almost evident corollary of~(\ref{6.0}). It can also be
considered as a special case of~(\ref{6.071}). In particular, to
commuting observables (in~$\Hil$) there correspond commuting bundle
morphisms (of $\HilB$):
	\begin{equation}	\label{6.074}
[\ope{A},\ope{B}]_{\_}=0   \iff   [\mor{A},\mor{B}]_{\_}=0.
	\end{equation}

	A little more general is the result, following
from~(\ref{6.073}), that to observables whose commutator is a c-number
there correspond bundle morphisms with the same c-number as a
commutator:
	\begin{equation}	\label{6.075}
[\ope{A},\ope{B}]_{\_}=c \, (\id_\Hil)   \iff   [\mor{A},\mor{B}]_{\_} =
							c \, (\id_\bHil).
	\end{equation}
for some $c\in\mathbb{C}$. In particular, the bundle analogue of the
famous relation $[\ope{Q},\ope{P}]_{\_}=\ih \, (\id_\Hil)$
between a coordinate $\ope{Q}$ and the conjugated to it momentum
$\ope{P}$ is $[\mor{Q},\mor{P}]_{\_}=\ih \, (\id_\bHil)$.

	A bit more complicated is the case for operators and morphisms along
paths at different `moments'.
Let $\gamma\colon J\to\base$ and $r,s,t\in J$.
If
$\Breve{\ope{G}}_{s,t}\colon(\ope{A},\ope{B})\mapsto
\ope{G}(\ope{A}(s),\ope{B}(t))$,
we define the bundle analogue $\Breve{\mor{G}}_{s,t}$ of
$\Breve{\ope{G}}_{s,t}$ by
\[
\Breve{\mor{G}}_{s,t}\colon
(\mor{A},\mor{B})\mapsto \Breve{\mor{G}}_{s,t}(\mor{A},\mor{B})\colon
\gamma \mapsto\Breve{\mor{G}}_{\gamma;s,t}(\mor{A},\mor{B})\colon
\proj^{-1}(\gamma(J)) \to\proj^{-1}(\gamma(J)),
\]
where
	\begin{multline} 	\label{6.081}
%	\begin{equation}	\label{6.081}
%	\begin{split}
\left.{\Breve{\mor{G}}}_{\gamma;s,t}
(\mor{A},\mor{B})\right|_{\bHil_{\gamma(t)}}
:=
l_{\gamma(r)}^{-1}\circ \ope{G}(\ope{A}(s),\ope{B}(t)) \circ l_{\gamma(r)}
\\
=
l_{\gamma(r)}^{-1}\circ
\ope{G}(
l_{\gamma(r)}\circ\Breve{\ope{A}}_{\gamma;s}(r)\circ l_{\gamma(r)}^{-1} ,
l_{\gamma(r)}\circ\Breve{\ope{B}}_{\gamma;t}(r)\circ l_{\gamma(r)}^{-1}
)
\circ l_{\gamma(r)}
\colon \bHil_{\gamma(r)} \to \bHil_{\gamma(r)}. %\qquad r\in J.
%	\end{split}
%	\end{equation}
	\end{multline}
Here
	\begin{equation}	\label{6.082}
\Breve{\ope{A}}_{\gamma;t}(r) :=
l_{\gamma(r)}^{-1}\circ\ope{A}(t)\circ l_{\gamma(r)} =
l_{t\to r}^{\gamma}\circ\mor{A}(t)\circ l_{r\to t}^{\gamma}
\colon \bHil_{\gamma(r)} \to \bHil_{\gamma(r)},
	\end{equation}
where~\eqref{6.0} has been used and
 $l_{s\to t}^{\gamma}:=l_{\gamma(s)\to\gamma(t)}$ is the (flat) linear
transport (along paths) from $\gamma(s)$ to $\gamma(t)$ assigned to the
isomorphisms $l_x$, $x\in\base$ (see Sect.~\ref{new-I},
equation~\eqref{4.12h}).%
\footnote{%
According to~\cite[sections~2 and~3]{bp-TP-morphisms} the morphism
 $\Breve{\mor{A}}_{\gamma;t}(r)$ along $\gamma$
is obtained via linear transportation of $\mor{A}_\gamma(t)$ along $\gamma$ by
means of the induced by $l_{s\to t}^{\gamma}$ linear transport along paths in
the fibre bundle  $\morf{\HilB}$ of bundle morphisms of $\HilB$%
% (see also subsection~\ref{VII.1}).%
}
Now the analogue of~\eqref{6.072} is
	\begin{multline}	\label{6.083}
\Breve{\ope{A}}_{1;\gamma;t_1}(r)\circ
\Breve{\ope{A}}_{2;\gamma;t_2}(r)\circ \cdots \circ
\Breve{\ope{A}}_{k;\gamma;t_k}(r)
\\
=
l_{\gamma(r)}^{-1}\circ
( \ope{A}_1(t_1)\circ \ope{A}_2(t_2)\circ\cdots\circ \ope{A}_k(t_k)\circ )
\circ l_{\gamma(r)}.
	\end{multline}
So, if $\ope{G}$ is a polynom or a convergent power series, the morphism
$\Breve{\mor{G}}_{\gamma;s,t}(\mor{A},\mor{B})$
along  $\gamma$ depends only on
 $\Breve{\mor{A}}_{\gamma;s}(r)$ and $\Breve{\mor{B}}_{\gamma;t}(r)$.

	In particular for $\ope{G}(\cdot,\cdot)=[\cdot,\cdot]_{\_}$, have
	\begin{equation}\label{6.084}
\left[ \Breve{\mor{A}}_{\gamma;s}(r) ,
		\Breve{\mor{B}}_{\gamma;t}(r) \right]_{\_}
=
l_{\gamma(r)}^{-1}\circ [\ope{A}(s),\ope{B}(t)]_{\_}\circ l_{\gamma(r)}
	\end{equation}
which for $s=r=t$ reduces to~(\ref{6.073}). In this case the analogues
of~(\ref{6.074}) and~(\ref{6.075}) are
	\begin{align}
	\label{6.085}
 [\ope{A}(s),\ope{B}(t)]_{\_} =0 &\iff
\left[
\Breve{\mor{A}}_{\gamma;s}(r) , \Breve{\mor{B}}_{\gamma;t}(r)
\right]_{\_} =0,
\\
	\label{6.086}
 [\ope{A}(s),\ope{B}(t)]_{\_} =c \, (\id_\Hil) &\iff
\left[
\Breve{\mor{A}}_{\gamma;s}(r) , \Breve{\mor{B}}_{\gamma;t}(r)
\right]_{\_} = c \, (\id_{\bHil_{\gamma(r)}}),
	\end{align}
respectively.

	The above considerations can \emph{mutatis mutandis}, e.g by
replacing $\gamma(t)$ with $x$,  $\ope{A}(t)$ with $\ope{A}$,  $\mor{A}$ with
$\overline{\mor{A}}$, etc., be transferred to global morphisms of $\HilB$, but
this is not needed for the present investigation.

	Further we will need a kind of `extension' of the
differentiation along paths $\mor{D}_{t}^{\gamma}$ (see~(\ref{5.7})) on some
fibre morphisms that will be presented below.

	Let $\Morf^p\HilB$, $p\in\mathbb{N}\cup{0}$,
 be the set of $C^p$ bundle morphisms of
$\HilB$. We define
$\tilde \mor{D}_{t}^{\gamma}$ to be the
\emph{differentiation along $\gamma$ (at $t$) of bundle morphisms}
from
\( \Morf^1\HilB \)
linearly acting on state vectors, mapping them on
\( \left.\Morf^0\HilB\right|_{\gamma(J)} \)
and given by
	\begin{equation}	\label{6.3}
\tilde \mor{D}_{t}^{\gamma}\colon  C \mapsto \tilde \mor{D}_{t}^{\gamma}(C)
:=
\mor{D}_{t}^{\gamma}\circ\left.C\right|_{\gamma(t)}
	\end{equation}
where $C\in\Morf^1\HilB$
\emph{%
acts on state vectors} (only).

	Applying~(\ref{5.8}), we can find the explicit (matrix) action of
\(\tilde \mor{D}_{t}^{\gamma}\).
	Let
\( C_t:=C|_{\gamma(t)} \) and
\( \pmb{\boldsymbol{[}} X \pmb{\boldsymbol{]}} \)
be the matrix of a vector or an operator $X$ in $\{e_a\}$.
	Due to~(\ref{5.8}), we have
\(
\pmb{\boldsymbol{[}}
	\tilde \mor{D}_{t}^{\gamma}(C) \Psi_\gamma(t) \pmb{\boldsymbol{]}} =
\left(
\frac{d}{dt} \boldsymbol{C}_t
\right)
\boldsymbol{\Psi}_\gamma(t) +
\boldsymbol{C}_t
\left( \frac{d}{dt} \boldsymbol{\Psi}_\gamma(t) \right) +
\boldsymbol{\Gamma}_\gamma(t)
\boldsymbol{C}_t
\boldsymbol{\Psi}_\gamma(t).
\)
	Substituting here \( \frac{d}{dt} \boldsymbol{\Psi}_\gamma(t) \)
from~(\ref{5.1}) and using~(\ref{5.10}), we obtain the matrix equation
	\begin{equation}	\label{6.4}
\pmb{\boldsymbol{[}}
	\tilde \mor{D}_{t}^{\gamma}(C) \Psi_\gamma(t) \pmb{\boldsymbol{]}} =
\left( \frac{d}{dt} \boldsymbol{C}_t \right)
\boldsymbol{\Psi}_\gamma(t) +
\left[
\boldsymbol{\Gamma}_\gamma(t),\boldsymbol{C}_t
\right]_{\_}
\boldsymbol{\Psi}_\gamma(t),
	\end{equation}
where $[\cdot,\cdot]_{\_}$ is the commutator
of the corresponding matrices, or
	\begin{equation}	\label{6.5}
\pmb{\boldsymbol{[}} \tilde \mor{D}_{t}^{\gamma} (C) \pmb{\boldsymbol{]}} =
\frac{d}{dt} \boldsymbol{C}_t +
\left[
\boldsymbol{\Gamma}_\gamma(t),\boldsymbol{C}_t
\right]_{\_} .
	\end{equation}

	Nevertheless that the last equation is valid in any local basis
it cannot be written in an invariant (operator) form  as the action of
$\frac{d}{dt}$ on bundle morphisms or sections is not defined, as well
as to $\boldsymbol{\Gamma}_\gamma(t)$ alone there does not correspond
some invariant operator or morphism.

	We derived~(\ref{6.5}) under the assumption that
\( \tilde \mor{D}_{t}^{\gamma} \)
acts on state vectors, i.e on ones satisfying the matrix-bundle Schr\"odinger
equation~(\ref{5.1}). Conversely, if we apply~(\ref{6.5}) to some vector
\( \Phi_\gamma(t)\in \bHil_{\gamma(t)} \) and compare the result with the one for
\( \left( \mor{D}_{t}^{\gamma}(C) \right) ( \Phi_\gamma(t) ) \)
obtained through~(\ref{5.8}) (see above), we see that \(\Phi_\gamma(t)\)
satisfies~(\ref{5.1}). Consequently, equation~(\ref{6.5}) is valid if
and only if it is applied on vectors representing the evolution of a
quantum system. Hence \(\Psi_\gamma(t)\) is a state vector, i.e.\  it
satisfies, for instance,
the bundle Schr\"odinger equation~(\ref{5.13}), iff in any basis the
equation~(\ref{6.4}) is valid for any bundle morphism $C$.
	In particular~(\ref{6.4}) is valid for the (Hermitian) morphisms
(along paths) corresponding to observables and \(\Psi_\gamma(t)\)
satisfying the bundle Schr\"odinger equation~(\ref{5.13}).

	The over-all above discussion shows the equivalence of~(\ref{6.4})
(for every morphisms $C$) with the Schr\"odinger equation (in anyone of its
(equivalent) forms mentioned until now). That is why~(\ref{6.4}) can be
called
\emph{%
matrix-morphism Schr\"odinger equation.}

\section{Conclusion}
\label{conclusion-II}

	Here we have continued to apply the fibre bundle formalism to
nonrelativistic quantum mechanics. We derived different forms of the bundle
Schr\"odinger equation which governs the time evolution of state sections
along paths in the Hilbert bundle description.

	In this description, as we have seen, the observables are described
via Hermitian bundle morphisms along paths. We also have concerned some
technical problems connected with functions of observables.

	In the future continuation of the present series we plan to consider
from a fibre bundle point of view the following items: pictures and integrals
of motion, mixed states, evolution transport's curvature, interpretation of
the theory and its possible further developments.

%>>>>>>>>>>>>>>>>>>>>>>>>>>>>>>>>>>>>>>>>>>>>>>>>>>>>>>>>>>>>>>>>>>>>>>
%<<<<<<<<<<<<<<<<<<<<<<<<<<<<<<<<<<<<<<<<<<<<<<<<<<<<<<<<<<<<<<<<<<<<<<

\addcontentsline{toc}{section}{References}

\bibliography{bozhopub,bozhoref}
\bibliographystyle{unsrt}

\addcontentsline{toc}{subsubsection}{\vspace{1ex}This article ends at page}

\end{document}